\documentclass[aps,pra,twocolumn,showpacs,superscriptaddress]{revtex4}
\usepackage{dcolumn}
\usepackage{bm}
\usepackage{graphicx}
\usepackage{amsmath}
\usepackage{latexsym}
\usepackage{amsfonts}
\usepackage{amssymb}
\usepackage{array}
\usepackage{epsfig}
\usepackage{times}
\usepackage{subfigure}

\newcommand{\ket}[1]{\left\vert#1\right\rangle}

\newcommand{\sprod}[2]{\left\langle#1\vert#2\right\rangle}

\newcommand{\bra}[1]{\left\langle#1\right\vert}

\newcommand{\opa}[1]{\hat{a}_{#1}}

\newcommand{\pare}[1]{\left(#1\right)}
\newcommand{\be}{\begin{equation}}
\newcommand{\ee}{\end{equation}}
\newcommand{\ba}{\begin{array}}
\newcommand{\ea}{\end{array}}
\newcommand{\beqn}{\begin{eqnarray}}
\newcommand{\eeqn}{\end{eqnarray}}

\begin{document}
\title{Generation of graph-state streams}

\author{Daniel Ballester}
\affiliation{School of Mathematics and Physics, Queen's University, Belfast BT7 1NN, United Kingdom}
\affiliation{Departamento de Qu\'imica F\'isica, Universidad del Pa\'is Vasco-Euskal Herriko Unibertsitatea, Apdo. 644, 48080 Bilbao, Spain}

\author{Jaeyoon Cho}
\author{M. S. Kim}
\affiliation{Institute for Mathematical Sciences, Imperial College London, London SW7 2PG, United Kingdom}
\affiliation{Optics Section, Blackett Laboratory, Imperial College London, London SW7 2BW, United Kingdom}

\date{\today}

\begin{abstract}
We propose a protocol to generate a stream of mobile qubits in a graph state through a single stationary parent qubit and discuss two types of its physical implementation, namely, the generation of photonic graph states through an atom-like qubit and those of flying atoms through a cavity-mode photonic qubit. The generated graph states fall into an important class that can hugely reduce the resource requirement of fault-tolerant linear optics quantum computation, which was previously known to be far from realistic. In regard to the flying atoms, we also propose a heralded generation scheme, which allows for high-fidelity graph states even under the photon loss.
\end{abstract}


\maketitle


The graph state, marked by having peculiar correlations \cite{bri01} and being a universal resource for quantum computation \cite{rau01}, has been recognized as one of the most important classes of many-body entangled states. In the last decade, numerous studies have been conducted to generate graph states in various systems \cite{sch05,cho05,lee08,an09,met07,lin09,bro05,wal05,pre07,che07,tam07} and the proof-of-principle experiments have been performed with realizing measurement-based logic operations using graph states \cite{wal05,pre07,che07,tam07}.

An important feature of the graph state is that it can be represented by a graph with each vertex representing a qubit. For a given graph (specifically, a simple graph \cite{wil79}), the corresponding graph state is defined as the state generated by preparing every qubit in state $\ket{+}=\frac{1}{\sqrt{2}}(\ket{0}+\ket{1})$, where $\ket{0}$ and $\ket{1}$ are the computational basis states, and subsequently performing a CPHASE operation between every pair of qubits connected by an edge. This definition is very suggestive in that one can essentially {\em build up} graph states in a systematic way, i.e., by adding qubits one by one, connecting two graph states, and so forth. This building-up procedure can be performed even with nondeterministic CPHASE gates \cite{dua05}, leaving us a problem of finding an optimal strategy \cite{gro06}.

In most physical situations, however, such graph-state generations are largely restricted both spatially and temporally. {In particular, }when mobile qubits, e.g., photons or flying atoms, are of major concern, such restrictions are more crucial unless massive reroutings and clever spatial allocations are permitted even though the mobile-qubit graph states can be useful for various quantum information and communication protocols. It would thus be crucial towards useful quantum information processing to identify the capabilities of individual systems as a source of graph states.

In this paper, we propose a protocol to generate graph states of mobile qubits in a highly restricted but commonly faced situation wherein a linear stream of mobile qubits come out sequentially from a place containing a single stationary qubit, which we will call a parent qubit. In particular, we consider two types of systems: firstly, a stream of single photons generated recurrently by an atomic (or atom-like \cite{mic00}) qubit trapped in a cavity, and secondly, a stream of flying atoms passing sequentially through a cavity containing a single photon \cite{rai01}. For convenience, let us call the former System I and the latter System II. 

\begin{figure}
\begin{tabular}{cp{1.5em}c}
\multicolumn{3}{c}{\includegraphics[width=0.6\columnwidth]{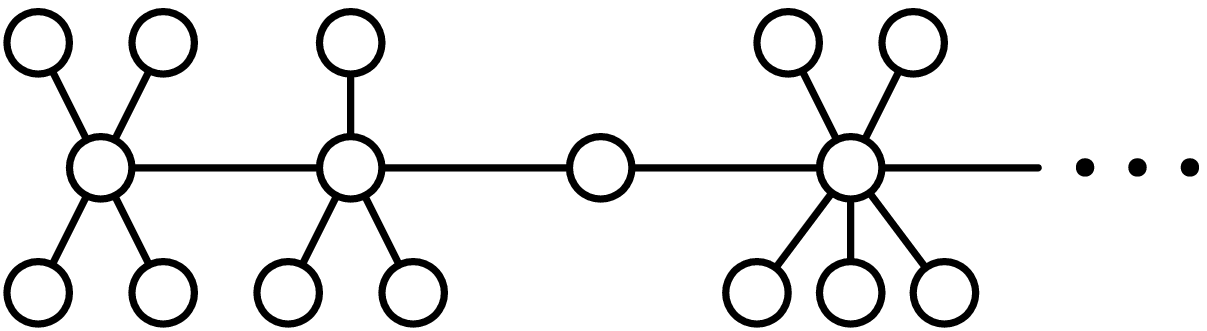}} \\
\multicolumn{3}{c}{(a) Linear chain of star graphs} \\[2.5ex]
\includegraphics[width=0.25\columnwidth]{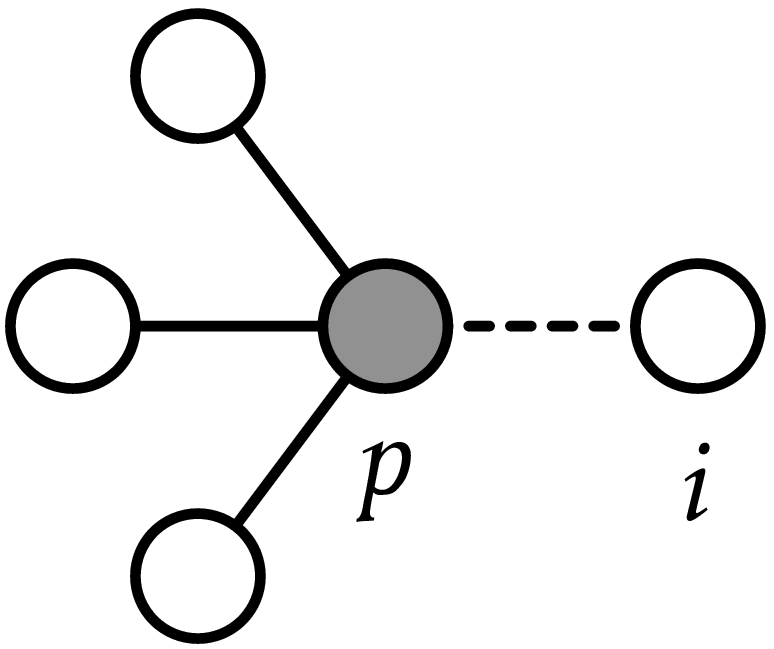} & &
\includegraphics[width=0.25\columnwidth]{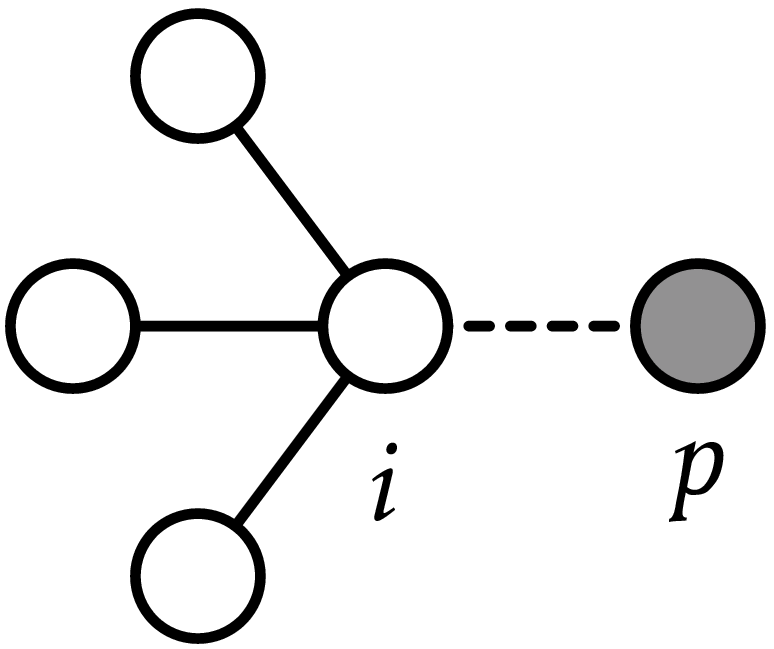} \\
(b) Branching & & (c) pulling-out
\end{tabular}
\caption{(a) An example of a graph state that can be generated by combining the two operations in (b) and (c). (b, c) Two unit operations for adding the $i$-th qubit. The shaded vertex represents the parent qubit. }
\label{fig1}
\end{figure}

Experimentally, only pairwise entanglement generation in System I and II using a parent qubit has been realized \cite{web09,rai01}.
While photonic graph-state generation in System I has been considered theoretically \cite{sch05,lin09}, the same for the atoms in System II has not been thoroughly studied. Our protocol offers a versatile and structured way of generating a rich variety of graph states and further envisions the possibility of the generation in System II. It turns out that the graph states corresponding to a linear chain of arbitrary star graphs \cite{wil79}, as shown in \figurename~\ref{fig1}(a), can be generated. The graph states and variations considered in Ref.~\cite{lin09} for System I fall into the same class up to local unitary transformations, albeit not explicit therein. This type of graph states is indeed of significant importance in fault-tolerant linear optics quantum computation (LOQC) \cite{daw06,cho07}. Although one can obtain fault-tolerance thresholds for LOQC, its requirement of resources---the number of two-photon entangled pairs, time steps, and parallel operations---is unrealistically high as it heavily relies on off-line preparation of particular graph states that is done with an extremely low success rate, spending a huge amount of the resources. Those graph states that the off-line preparation is aimed at are in fact of the same shape as that in \figurename~\ref{fig1}(a). Our scheme would thus hugely reduce the resource requirement of fault-tolerant LOQC. In regard to System II, experimental errors may be due to the loss of the photon that should be kept in the cavity throughout the whole process as a parent qubit. In order to circumvent this problem, we propose a heralding scheme that can single out successful events when the atoms are in high-fidelity graph states.  The schemes considered in this paper are within current technology.

Our protocol makes use of two unit operations to generate graph states. The first one is what we call a {\em branching} operation, which adds a vertex stemming from the parent qubit, as in \figurename~\ref{fig1}(b). This operation brings the parent qubit $p$ into a combination of the parent and a mobile qubit and maps the state of the parent qubit into the combined state as follows: $|0\rangle_p\rightarrow |0\rangle_p|+\rangle_i$ and $|1\rangle_p\rightarrow|1\rangle_p|-\rangle_i$, where $\ket{-}=\frac{1}{\sqrt{2}}(\ket{0}-\ket{1})$ and the subscript $i$ denotes the $i$-th mobile qubit.  This resembles a CPHASE operation between the parent and $i$-th mobile qubits prepared in the $|+\rangle$ state. The other operation is what we call a {\em pulling-out} operation, which adds a vertex taken by the parent qubit while a new mobile qubit takes the previous place of the parent qubit, as in \figurename~\ref{fig1}(c). This operation maps the state as $|0\rangle_p\rightarrow |0\rangle_i|+\rangle_p$ and $|1\rangle_p\rightarrow|1\rangle_i|-\rangle_p$, which is identical to a branching operation followed by a SWAP operation. While the general SWAP operation is a nonlocal operation, it is important to note that this particular SWAP operation can be replaced by local Hadamard operations. It is easily seen that by combining branching and pulling-out operations, one can generate a state corresponding to any linear chain of star graphs, as shown in \figurename~\ref{fig1}(a). 

\begin{figure}
\begin{tabular}{cp{1.5em}c}
\includegraphics[width=0.38\columnwidth]{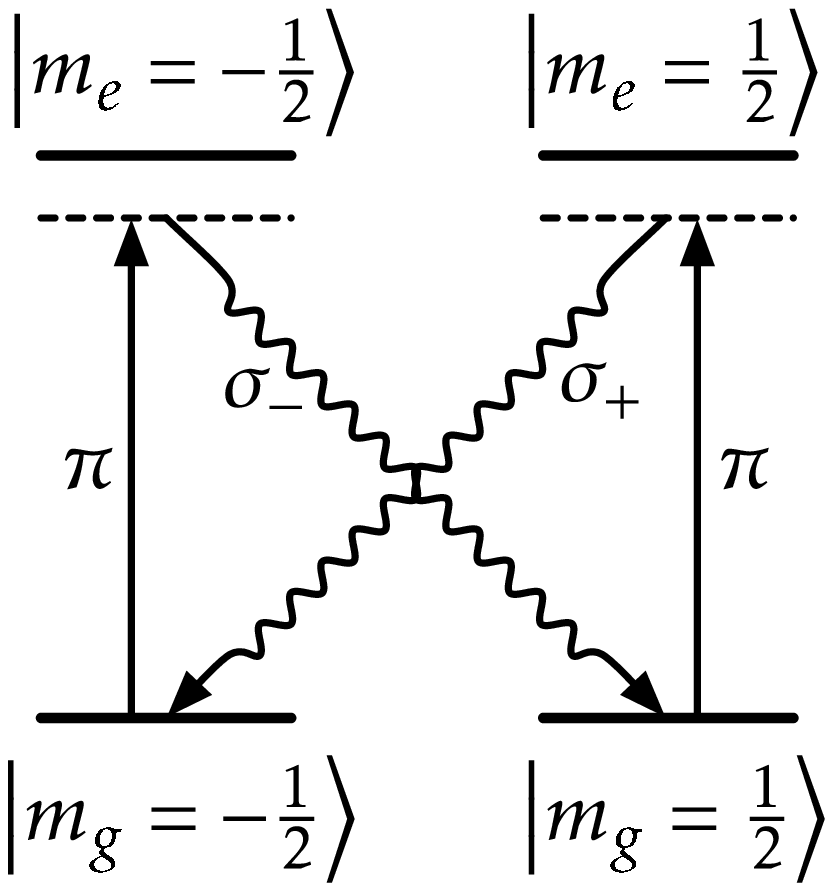} & &
\includegraphics[width=0.4\columnwidth]{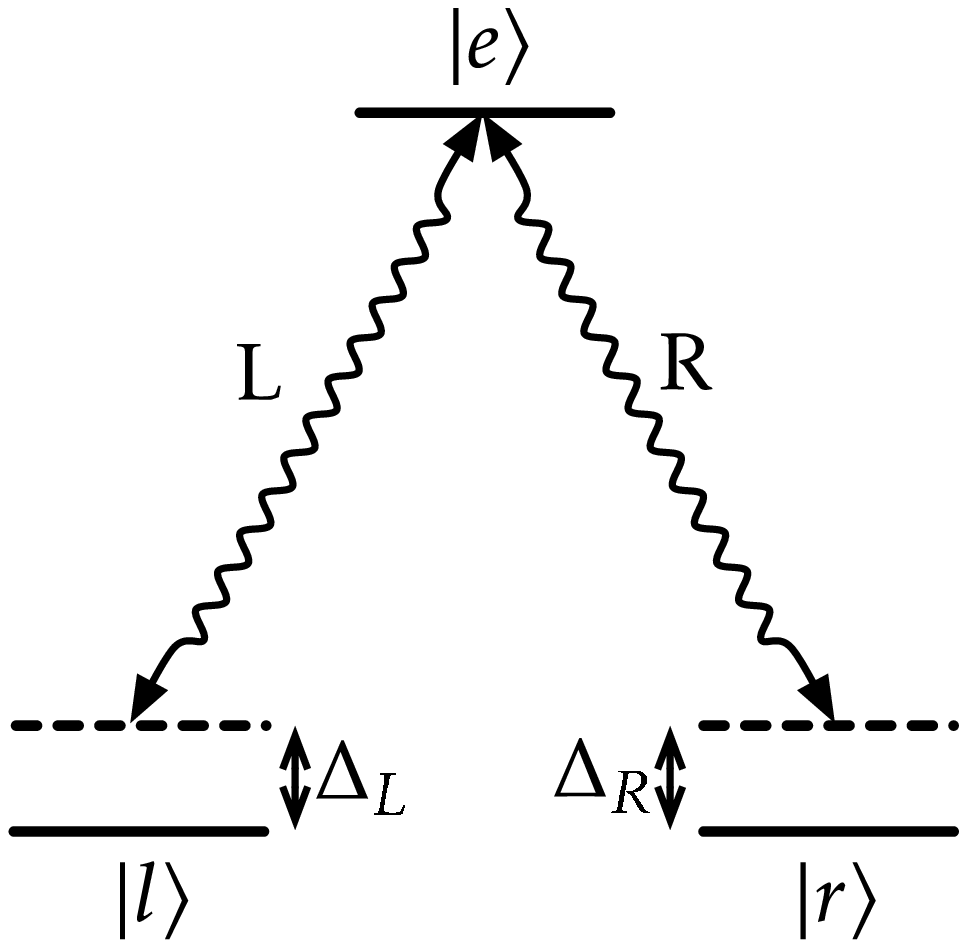} \\
(a) & & (b)
\end{tabular}
\caption{\label{fig2}Involved atomic levels and transitions for System I (a) and for System II (b). Two ground hyperfine levels represent a qubit.}
\end{figure}

In System I where photons are mobile qubits, single-qubit operations are tractable for both the parent and the mobile qubits. The essential part is thus the branching operation, as local Hadamard operations transform the branching operation into the pulling-out operation. For the branching operation, one needs to generate a single photon in such a way that an arbitrary superposition state of the parent qubit $\alpha\ket{0}_{p}+\beta\ket{1}_{p}$ is transformed to $\alpha\ket{0}_{s}\ket{\sigma_{+}}+\beta\ket{1}_{s}\ket{\sigma_{-}}$, where $\ket{\sigma_{\pm}}$ are two orthogonal polarization states of a photon. This can be done by using (or slightly modifying) well-studied existing schemes for single-photon generation \cite{kim08}. For example, one can realize the branching operation using two ground hyperfine levels $\ket{0}\equiv\ket{m_{g}=-\frac{1}{2}}$ and $\ket{1}\equiv\ket{m_{g}=\frac{1}{2}}$ with total spin $F=\frac{1}{2}$, as shown in \figurename~\ref{fig2}(a), to represent a parent qubit. As the transitions with $\Delta m=\pm 1$ are coupled to the $\sigma_\pm$-polarized modes of a cavity, Raman transitions can occur for both the ground levels by applying a $\pi$-polarized classical field, which generates a single cavity photon while flipping the atomic qubit. The single photon leaking out of the cavity then has a polarization depending on the atomic initial state, as we desired. The spin flip of the atomic state is easily correctable by a single-qubit operation, or this correction can be simply put off to a later time by employing the Pauli frame \cite{kni05}. The spontaneous emission can be suppressed by having a sufficiently large detuning. As for the time scale, using this conventional method of adiabatic transfer (see Ref. \cite{kim08}), the width of a single-photon pulse can be estimated as some large constant (say, $\sim$10) divided by the cavity decay rate, hence the repetition rate can be typically as high as tens of kHz.

Differently from System I,  realizing the branching and pulling-out operations is not straightforward in System II as it is hard to perform single-qubit operations on the parent qubit, which is encoded in a cavity-mode photon. In this case, the CPHASE and SWAP operations should be directly implemented. For this, we consider a $\Lambda$-type level structure as shown in \figurename~\ref{fig2}(b). The two transitions are coupled, respectively, to the two orthogonally polarized modes of the cavity, but are largely detuned from the resonance. Consequently, when the atom just passes through, no atom-cavity interaction takes place. In order to turn on the interaction, we control the two detunings separately by means of an ac Stark shift. As will be shown later, when only one transition is brought into resonance, a CPHASE operation can be performed, and when both transitions are brought into resonance, a SWAP operation can be performed. We can thus perform these operations selectively while the flying atom is passing through the interaction region of the cavity by applying timely pulses that induce the ac Stark shifts. Note that as the flying atoms enter the cavity with a random timing in experiments, the existence of an atom should be detected optically for a correct timing of the pulses. Although this protocol would benefit from the longer coherence time achievable in state-of-the-art microwave cavity QED technologies \cite{rai01}, there are still practical shortcomings of them that make our discussion for System II more relevant to optical cavities \cite{kim06, wil07} at the present time. For example, while atoms with the required level structure are readily available in the optical regime, this is not the case in the microwave regime (i.e., Rydberg atoms). 

Whereas the essential component of System I---coherent generation of a single photon by a Raman transition---has been well established \cite{kim08}, System II needs further analysis, which we focus on in the remainder of this paper. As shown in \figurename~\ref{fig2}(b), the atom has two degenerate ground levels $\ket{l}$ and $\ket{r}$, which represent a qubit, and an excited level $\ket{e}$. The transition $\ket{l}\leftrightarrow\ket{e}$ ($\ket{r}\leftrightarrow\ket{e}$) is coupled to the $L$($R$)-polarized mode of the cavity with coupling strength $g_{L}$ ($g_{R}$) and detuning $\Delta_{L}$ ($\Delta_{R}$). In case $\Delta_{L}=\Delta_{R}=0$, the interaction Hamiltonian can be written as
\begin{equation}
\mathcal{H}_{\text{I}}=\sum_{\mu=L,R}g_{\mu}(a_{\mu}\sigma_{\mu}^{\dagger}+a_{\mu}^\dag\sigma_{\mu}),
\label{ham1}
\end{equation}
where $a_{\mu}$ denotes the annihilation operator for the $\mu$-polarized mode of the cavity, $\sigma_{L}^{\dagger}=\ket{e}\bra{l}$, and $\sigma_{R}^{\dagger}=\ket{e}\bra{r}$. We assume $g_{L}=g_{R}=g$ when the transitions are resonant. If the atomic transition is largely detuned ($\Delta_{\mu}\gg g_{\mu}$), the effective coupling between the atom and the external field can be regarded as being zero, i.e., $g_{L}=g_{R}=0$. 

{We first outline our scheme in an idealized situation where both the atomic and the cavity decays are absent.} At the beginning, the cavity should be loaded with a single photon. This can be done in combination with an entangling operation. For this, the first atom is initially prepared in state $\ket{e}$ and a classical pulse that induces the ac Stark shift is applied to adjust the detunings to achieve $\Delta_{L}=\Delta_{R}=0$. After straightforward calculation based on the Hamiltonian  $\mathcal{H}_{\text{I}}$ we can find that if the pulse duration $\tau_{0}$ is chosen to be $g\tau_{0}=\pi/2\sqrt{2}$, the state is transformed to $\frac{1}{\sqrt{2}}(\ket{l}\ket{L}+\ket{r}\ket{R})$, which is the two-qubit graph state.

From the second atom, either a branching or a pulling-out operation is performed as outlined before. In order to perform a CPHASE operation, only $\Delta_{R}$ is adjusted to be zero while {$\Delta_{L}\gg g_L$} during a period of time $\tau_{1}$ such that $g_{R}\tau_{1}=\pi$. This operation transforms the state as $\ket{r}\ket{R}\rightarrow-\ket{r}\ket{R}$ while leaving $\ket{l}\ket{L}$, $\ket{l}\ket{R}$, and $\ket{r}\ket{L}$ unchanged, which is identical to a CPHASE operation. In order to perform a SWAP operation, we adjust both $\Delta_{L}$ and $\Delta_{R}$ to be zero during a period of time $\tau_{2}$ such that $g\tau_{2}=\pi/\sqrt{2}$. Solving the Schr\"{o}dinger equation for the Hamiltonian  (\ref{ham1}), it can be shown that this operation transforms the state as follows: $\ket{l}\ket{L}\rightarrow-\ket{r}\ket{R}$, $\ket{l}\ket{R}\rightarrow\ket{l}\ket{R}$, $\ket{r}\ket{L}\rightarrow\ket{r}\ket{L}$, and $\ket{r}\ket{R}\rightarrow-\ket{l}\ket{L}$.  This is identical to a SWAP operation followed by a ZX operation acted on both qubits, where $Z$ and $X$ are the Pauli operators. The correction of these Pauli operations can be deferred by employing the Pauli frame \cite{kni05}.

\begin{figure}
\centering{
\includegraphics[width=0.3\textwidth]{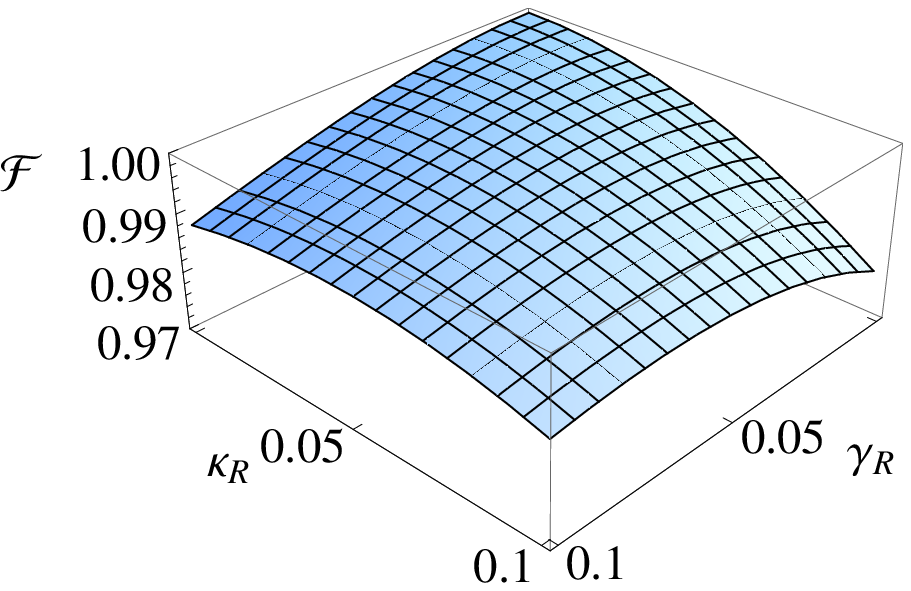} \\
(a) \\[2.5ex]
\includegraphics[width=0.3\textwidth]{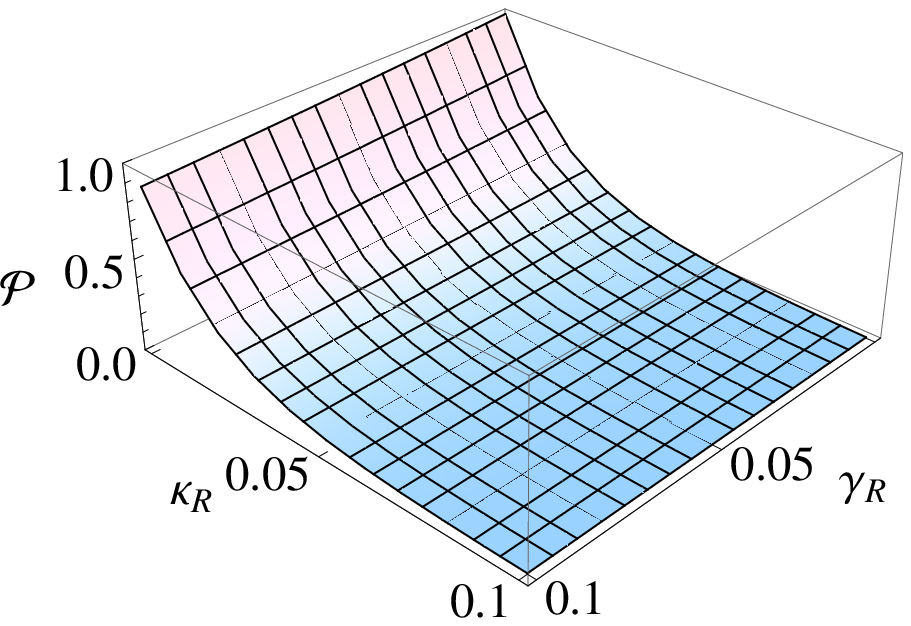} \\
(b)\\[2.5ex]
\includegraphics[width=0.3\textwidth]{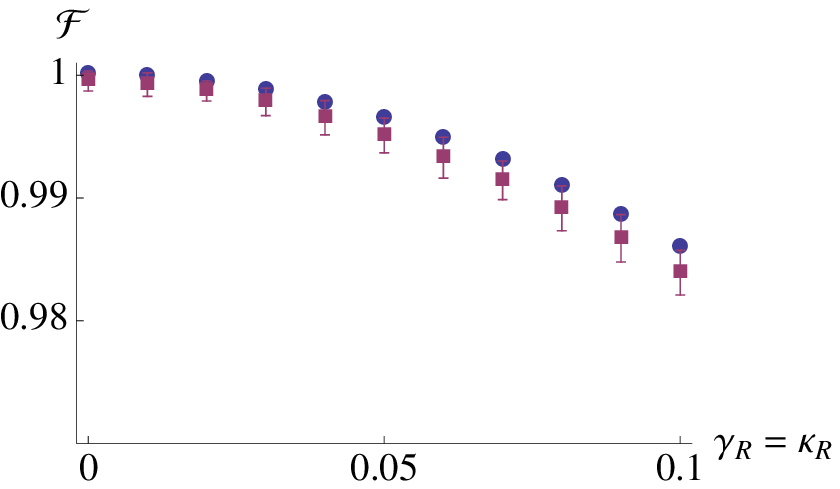} \\
(c) \\[2.5ex]
\includegraphics[width=0.3\textwidth]{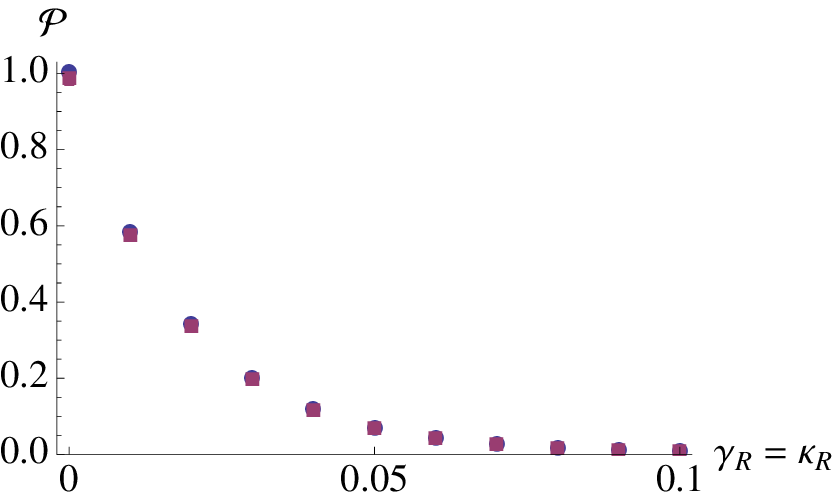} \\
(d)
}
\caption{\label{fig3} (a) Fidelity and (b) probability of the heralded generation of a 3-qubit cluster state versus the cavity damping and spontaneous emission rates. (c) Fidelity and (d) probability of the heralded generation of the same state versus the cavity damping (here $\gamma_R=\kappa_R$) in the protocol described (circles), and assuming that there is up to a 10\% uncertainty (uniformly sampled) in the selection of the interaction time (squares). We take $\kappa_L=\kappa_R$ and $\gamma_L=\gamma_R$ in all cases. Cavity damping and spontaneous emission rates are measured in units of the coupling constant, $g$, which is set equal to unity.}
\end{figure}

In real experiments, the above idealized situation is not possible to achieve due to the finite finesse of the resonator, together with the finite lifetime for the excited state of the three-level atoms. To model the evolution of the chain of atoms as they cross the cavity, we can use a Lindblad equation under the Markovian approximation for the density matrix of the system \cite{barnett97}
\begin{eqnarray}
\frac{\partial \varrho}{\partial t} = -i [{\cal H}, \varrho] - \sum_{\mu=L,R}\kappa_\mu \pare{ \opa{\mu}^\dag \opa{\mu} \varrho + \varrho \opa{\mu}^\dag \opa{\mu} - 2 \opa{\mu} \varrho \opa{\mu}^\dag }&& \nonumber \\
- \sum_{\mu=L,R}\gamma_\mu \pare{ \hat{\sigma}_{\mu}^\dag  \hat{\sigma}_{\mu} \varrho + \varrho  \hat{\sigma}_{\mu}^\dag  \hat{\sigma}_{\mu}- 2  \hat{\sigma}_{\mu} \varrho  \hat{\sigma}_{\mu}^\dag } . && \label{master}
\end{eqnarray}
Here $\kappa_\mu$ and $\gamma_\mu$ stand for the cavity damping and {the atomic} spontaneous emission rates {for} the 
$\mu-$circularly polarized mode, respectively. In writing the master equation (\ref{master}) we have assumed that the cavity field is coupled to a bosonic environment at zero temperature. 

Clearly, the performance of our protocol depends on the ability of the resonator to keep the excitation for a time long enough to perform all the operations needed. This means that the fidelity of the state generated will be very much affected by the damping and spontaneous emission rates. 
In spite of the photon loss, however, the fidelity can be drastically improved nearly as high as unity by adding to the original scheme a heralding process based upon the detection of the photon leaking out of the cavity. Note that the number of excitations $\sum_{\mu=L,R}a_{\mu}^{\dagger}a_{\mu}+\sum_{i}\ket{e}_{i}\bra{e}$ in an ideal case is always kept to be one, while the environmental effect only decreases it. Consequently, once a photon is detected, it is guaranteed that all the preceded operations have been performed along with a single excitation kept in the system, and hence the fidelity of the final state should be high. We will take the state only when the photodetector clicks (in the computational basis) after the complete set of gates have been applied to the chain of atoms. 

Another important advantage of employing the heralding process is that we can deduce the fidelity and the success probability for an arbitrary number of atoms in a pseudo-analytic way, avoiding to calculate the whole evolution for all the elements of the huge density matrix. For convenience, let us change the notation to account explicitly for the number of photons in the cavity for each polarization, i.e. $\ket{10} \equiv \ket{L}$, $\ket{01} \equiv \ket{R}$, while $\ket{00}$ will denote the absence of the photon in any polarization. To calculate the evolution of one atom together with the cavity field (within the one or zero-excitation subspace), we can expand the density matrix using the basis $\{ \ket{l10}, \ket{l01}, \ket{r10}, \ket{r01}, \ket{e00}, \ket{l00}, \ket{r00}  \}$ to solve the master equation (\ref{master}). After the ideal evolution for a CPHASE and a SWAP, only the $4\times4$ sub-matrix at the top left corner for those states containing one photon will have non-zero elements. If the heralding process is assumed, only these elements indeed make contributions to the final state. It turns out that even with the photon loss, this sub-matrix, when renormalized, remains as a pure state, which can be written as $\alpha\ket{10}\ket{\phi_{L}}+\beta\ket{01}\ket{\phi_{R}}$, where $\ket{\phi_{\mu}}$ is a normalized state representing other portion of the state. For the ensuing step, this pure state can be taken as the initial state, where the evolution can be calculated in the same manner as if a state $\alpha\ket{10}+\beta\ket{01}$ was the initial state. This is because the interaction exists only between the cavity photon and the new atom, while all the earlier atoms left the cavity do not take part in the new evolution.

As a particular example, we have performed numerical calculations for the generation of a 3-qubit graph state. In the ideal case this linear graph state $\ket{\phi^{id}_3}$ will be equivalent to a 3-qubit GHZ state. Fig. \ref{fig3}{(a)} shows the fidelity of the state obtained after heralding and the probability for this to happen with respect to the cavity damping rate and the atomic spontaneous emission rate. As we are comparing pure states, $\ket{\phi_3}$ and $\ket{\phi^{id}_3}$, the fidelity can be calculated as ${\cal F} = \left| \sprod{\phi_3}{\phi^{id}_3} \right|^2$. Typical values of $(g,\gamma,\kappa)/2\pi$ are around $ (16,3,1.25)$ MHz for state-of-the-art experiments with Rb atoms \cite{rem09} or $ (34,2.6,4.1)$ MHz for Cs \cite{kim06}. Taking these parameters, the fidelities are ${\cal F}\approx 0.886$ and ${\cal F}\approx 0.998$, respectively, while the probabilities of heralding are ${\cal P}\approx 0.84\%$ and ${\cal P}\approx 0.19\%$. This is relevant in order to calculate the repetition rate which can be obtained with this protocol. In the evolution we have also {taken into account} an extra idle time of magnitude $g\tau_{\rm idle}=\pi$ after every atom crosses the cavity. This might be necessary to ensure the right spacing between them. As a result, one 3-qubit graph state is yielded every ~30.2 $\mu$s and ~62.6 $\mu$s on average, respectively. We have also studied the resilience of our protocol versus a hypothetical fluctuation in the interaction time. {This also models experimental imperfections in selecting atomic velocities and controlling the detunings.} Figs. \ref{fig3} (c) and (d) show the decrease in fidelity and probability of heralding, respectively, assuming that there is up to a 10\% uncertainty (uniformly sampled) in the selection of the interaction time. {The numerical results indicate that our scheme is} robust against this type of imperfection. 

As a final remark, the most crucial aspect to judge if the generation of the graph state has been successful after all is to determine whether the state $\ket{\phi_3}$ is really entangled or not. In spite of the lack of general criteria for multipartite entanglement, we can still make use of the entanglement witness derived in Ref.  \cite{tot08,nah}. In our case, this particular type of witness for graph states allows to relate the existence of entanglement in the state with its fidelity. More precisely, the witness says that any state close to a graph state with fidelity larger than 1/2 should be entangled.

The authors thank T. Rudolph and H. Nha for valuable discussions and insights. Funding from ESF and UK EPSRC is gratefully acknowledged.


\bibliography{references}{}

\end{document}